\documentclass[12pt]{article}
\usepackage{epsfig}

\def\beq{\begin{equation}}
\def\eeq#1{\label{#1}\end{equation}}
\def\eeqn{\end{equation}}

\def\beqa{\begin{eqnarray}}
\def\eeqa#1{\label{#1}\end{eqnarray}}
\def\eeqan{\end{eqnarray}}

\let\bar=\overbar

\def\Dslash{\not{\hbox{\kern-4pt $D$}}}
\def\dslash{\not{\hbox{\kern-2pt $\del$}}}

\def\msb{{\bar{\ssstyle M \kern -1pt S}}}

\def\Title#1{\begin{center} {\Large {\bf #1} } \end{center}}

\begin{document}

\Title{Prospects of mixing and CP violation in the D system at $SuperB$}

\bigskip\bigskip


\begin{raggedright}  

{\it Paolo Branchini\index{Branchini, P.}\\
INFN Roma Tre \\
Via Della Vasca Navale 84\\
00146 Rome, ITALY}
\bigskip\bigskip
\end{raggedright}

\section{Abstract}

The $SuperB$ experiment hosted by the Cabibbo laboratory will have the possibility
to study electron positron collision both at the $\Upsilon(4S)$ center of mass and
at the $\psi(3770)$. Therefore the potential physics reach of the experiment will be greatly enriched. In this paper I present the implications such a variety
of measurements can have on various new physics scenarios.

\section{Introduction}

The absence of flavor changing neutral currents (FCNCs) was considered 
as an indication of the existence of a fourth quark. The GIM mechanism 
was proposed in this framework and within this framework explained the
suppression of some transitions \cite{GIM}. FCNCs still represent a very important 
issue in flavor physics when looking for new physics.
Another interesting aspect is that the Standard Model (SM) predicts 
small mixing and CP violation for charm mesons. While mixing has already been 
observed for $D^0$ mesons, CP violation in charm has still to be discovered and a time-dependent analysis may provide a tool to test the 
Cabibbo-Kobayashi-Maskawa mechanism \cite{CKM} and the SM itself through 
the measurement of the angle $\beta_{c,\mathrm{eff}}$ in the charm unitarity triangle.
This work describes the potential reach of $SuperB$ when studying rare
decays, mixing and CP violation in charm mesons. It is useful to stress here
the uniqueness of charm mesons in the understanding of the SM. $D^0$ mesons are the only mesons made from two \textit{up - type} quarks ($c$ $ \overline{u}$). It is
therefore necessary to study this system to improve our knowledge on the
flavor changing structure of the SM. Moreover the interpretation we have
nowadays of the $CP$ violating mechanism in terms of the CKM matrix
can be (dis)proved in the $up-sector$.

\section{Notations for flavor mixing and CPV}

Flavor mixing occurs when the Hamiltonian eigenstates  
$D_1$, $D_2$ differ from the flavor eigenstates $D^0$, $\overline{D}^0$. A neutral $D$ meson produced at time t = 0 in a definite flavor state $D^0$ will 
then evolve and oscillate into a state of opposite flavor $\overline{D}^0$ after a certain time. If we describe the time evolution of the
flavor eigenstates in terms of a 2x2 effective Hamiltonian, H = M - $i\over{2}$$ \Gamma$, then, assuming $CPT$ is conserved, the mass eigenstates can be expressed
 in terms of the flavor eigenstates by:
\begin{equation}
     |D_1\rangle=p|D^0\rangle+q|\overline{D}^0\rangle
\end{equation}
\begin{equation}
     |D_2\rangle=p|D^0\rangle-q|\overline{D}^0 \rangle
\end{equation}
with the normalization $|q|^2$ + $|p|^2$ = 1 and
\begin{equation}
    {q\over{p}}= \sqrt{{M^*_{12}-i/2 \Gamma^*_{12}}\over{M_{12}-i/2 \Gamma_{12}}}
\end{equation}
Assuming a phase convention such that $CP|D^0\rangle$ = -$|\overline{D}^0\rangle$ and 
$CP|\overline{D}^0\rangle$ = -$|D^0\rangle$ then,
if CP is conserved, we have that $q = p$ = $1/\sqrt{2}$ and 
the mass eigenstates coincide with the CP eigenstates: 
$|D_1\rangle$ = $|D_{CP-}\rangle$ $(CP-odd)$ and $|D_2\rangle$ = $|D_{CP+}\rangle$ $(CP-even)$.
The mixing parameters x, y, can be expressed in terms of the difference 
of masses $(m_{1,2})$ and widths ($\Gamma_{1,2}$) of the Hamiltonian eigenstates,
\begin{equation}
   x={{m_2-m_1}\over{\Gamma}};y={{\Gamma_2-\Gamma_1}\over{2\Gamma}}
\end{equation}
where $\Gamma$=$(\Gamma_1+\Gamma_2)/2$.
$CP$ violation can be of three types:
\newline
1. $CPV$ $in$ $decay$ $or$ $direct$ $CPV$: this occurs when the decay amplitudes for CP
conjugate processes are different in modulus. If $\langle|H|D^0\rangle=A_f$,
$\langle \overline{f}|H|\overline{D}^0\rangle=\overline{A}_{\overline{f}}$ are the $D^0$ and
$\overline{D}^0$ decay amplitudes into the final $f$ and CP conjugate 
$\overline{f}$, then
\begin{equation}
   {\overline{A}_{\overline{f}}\over{A_f}} \neq 1 \rightarrow CPV
\end{equation}
\newline
2. $CPV$ $in$ $mixing$ $or$ $indirect$ $CPV$: it occurs when the 
Hamiltonian eigenstates do not coincide with the CP eigenstates. That is:
\begin{equation}
 |{q\over{p}}|\neq 1 \rightarrow CPV.
\end{equation}
\newline
3. $CPV$ $in$ $the$ $interference$ $of$ $mixing$ $and$ $decay$: 
for neutral $D$ mesons there is a third possibility to observe $CP$ violation even when $CP$ is conserved in mixing and also in decay. In this case,
$CP$ violation arises when, in a process with final state $f$ that can 
be reached by neutral $D$ mesons of both flavors (i.e. $D^0$ and
$\overline{D}^0$), there is a relative weak phase difference between the mixing and the decay
amplitudes. The quantity of interest that is independent of phase conventions,
and physically meaningful, is
\begin{equation}
\lambda_f \equiv {q\over{p}} {{\overline {A}_f}\over{A_f}}
\equiv |{q\over{p}}| |{{\overline{A}_f}\over{A_f}}|e^{i(\delta_f+\phi_f)}
\end{equation}
where $\delta_f$ and $\phi_f$ are the $CP$-conserving and 
$CP$-violating phases respectively.
If $CP$ is conserved in mixing and in decay, the signature of $CP$ 
violation in the interference of mixing and decay is thus
\begin{equation}
\sin(\phi_f) \neq0\rightarrow CPV.
\end{equation}
For $CP$ eigenstates, $CPV$ in either mixing or decay is indicated by
\begin{equation}
|\lambda_f|\neq 1,
\end{equation}
while $CPV$ in the interference of mixing and decay corresponds to
\begin{equation}
Im(\lambda_f)\neq 1.
\end{equation}
If there is no weak phase in the decay amplitudes then $arg (q/p)$ = $\phi$
and it is independent of the final state $f$.

\section{Experimental observables}
Time dependent analyses provide the most precise tests on the mixing parameters.
An excellent description of this approach is given in \cite{neri}.
In this paper I'll quote the results.
First of all when allowing for CPV to take place 10 parameters can be
extracted from the $\chi^2$ fit: $x$, $y$, $|q/p|$, $\phi$, $\delta$, 
$\delta_{k \pi \pi}$, $R_D$ $A_D$, $A_\pi$, $A_k$. The parameters $\delta$, $\delta_{k\pi\pi}$ are the relative strong phases, 
$R_D$ is the ratio $\Gamma(D^0\rightarrow K^+ \pi^-)$/$\Gamma(\overline{D}^0\rightarrow K^-\pi^+)$ and $A_D$, $A_\pi$, $A_K$ are the direct 
$CP$-violation asymmetries for the $\Gamma(D^0\rightarrow K^+ \pi^-)$,
 $\Gamma(D^0\rightarrow \pi^+ \pi^-)$ and  $\Gamma(D^0\rightarrow K^+ K^-)$
modes. The relationships between these parameters and the measured
observables are given below.

\begin{itemize}
\item $Semileptonic$ $decays$: they search for mixing by reconstructing the ''wrong-sign" (WS) decay chain, $D^{*+} \rightarrow D^0 \pi^+$, $D^0\rightarrow \overline{D}^0$, $\overline{D}^0\rightarrow K^{(*)+}e^- \overline{\nu_e}$. In contrast to
hadronic decays, the WS charge combinations can occur only through mixing.
The measurement of $R_M$ is related to the mixing parameters as follows
\begin{equation}
R_M= {1\over {2}}(x^2+y^2) 
\end{equation}
and can be obtained directly as the ratio of WS to the right sign (RS) signal
events. The RS events correspond to the non-mixed process.
\item $Decays$ $to$ $CP$ $eigenstates$: They measure the mixing parameters
$y_{CP}$ and the $CPV$ parameter $A_\Gamma$ with a lifetime ratio analysis of the transitions to the $CP$  eigenstates and the transitions to the CP-mixed state 
$D^0 \rightarrow K^-\pi^+$.
\begin{equation}
2 y_{CP} = (|{{q \over{p}}}|+|{p\over{q}}|)y cos \phi-(|{q \over{p}}|-|{p\over{q}}|)x  sin \phi
\end{equation}
\begin{equation}
2 A_{\Gamma} = (|{{q \over{p}}}|-|{p\over{q}}|)y  cos \phi-(|{q \over{p}}|+|{p\over{q}}|)x  sin \phi
\end{equation}
The paramater $A_\Gamma$ is the decay-rate asymmetry in $CP$-even eigenstates, e.g. in $D^0\rightarrow K^+K^-$ and $D^0\rightarrow \pi^+ \pi^-$, provides constraints on the mixing and $CPV$ parameters according to the relations:
\begin{equation}
{\Gamma(D^0\rightarrow K^+K^-)-\Gamma(\overline{D}^0\rightarrow K^- K^+)\over{\Gamma(D^0\rightarrow K^+K^-)+\Gamma(\overline{D}^0\rightarrow K^- K^+)}}=
A_K+{<t>\over{\tau_D}}A_{CP}^{indirect}
\end{equation}
\begin{equation}
{\Gamma(D^0\rightarrow \pi^+\pi^-)-\Gamma(\overline{D}^0\rightarrow \pi^- \pi^+)\over{\Gamma(D^0\rightarrow \pi^+\pi^-)+\Gamma(\overline{D}^0\rightarrow \pi^- \pi^+)}}=
A_\pi+{<t>\over{\tau_D}}A_{CP}^{indirect}
\end{equation}
and $A_{CP}^{indirect}$ is given by:
\begin{equation}
2 A_{CP}^{indirect} = (|{{q \over{p}}}|+|{p\over{q}}|)x  sin \phi-(|{q \over{p}}|-|{p\over{q}}|)y  cos \phi
\end{equation}
$<t>$ is the average reconstructed $D^0$ proper time and $\tau_D$ is the nominal
$D^0$ lifetime.
\item $Three-body$ $D^0\rightarrow K^0_s \pi^+\pi^-$ and $D^0\rightarrow K^0_s K^+K^-$ $decays$: They measure directly the mixing and $CPV$ parameters $x$, $y$, $|q/p|$ and $\phi$ with a time dependent Dalitz plot analysis.

\item $Wrong-sign$ $decays$ $to$ $hadronic$ $non-CP$ $eigenstates$: They measure the
parameters $x^{'\pm}$, and $y^{'\pm}$ and $R_D$ and $A_D$ in a time-dependent 
analysis of the WS events selected through the decay chain $D^{*+}\rightarrow D^0 \pi^+$, $D^0\rightarrow k^+ \pi^-$. The parameters are defined in the
following way:
\begin{equation}
    x^{'\pm}={{(1\pm A_M)\over{1\mp A_M}}}^{1/4}(x^{'} cos \phi \pm y^{'} sin \phi)
\end{equation}       
\begin{equation}
    y^{'\pm}={{(1\pm A_M)\over{1\mp A_M}}}^{1/4}(y^{'}  cos \phi \mp x^{'}  sin \phi)
\end{equation}       
\begin{equation}
{\Gamma(D^0\rightarrow K^+ \pi^-)+\Gamma(\overline{D}^0\rightarrow K^- \pi^+)\over{\Gamma(D^0\rightarrow K^-\pi^+)+\Gamma(\overline{D}^0\rightarrow K^+ \pi^-)}}= R_D
\end{equation}
\begin{equation}
{\Gamma(D^0\rightarrow K^+ \pi^-)-\Gamma(\overline{D}^0\rightarrow K^- \pi^+)\over{\Gamma(D^0\rightarrow K^-\pi^+)+\Gamma(\overline{D}^0\rightarrow K^+ \pi^-)}}= A_D
\end{equation}
and
\begin{equation}
 x^{'}=x  cos \delta + y  sin \delta
\end{equation}
\begin{equation}
 y^{'}=-x  sin \delta + y  cos \delta
\end{equation}
\begin{equation}
A_M = {{|q/p|^2-|p/q|^2}\over{|q/p|^2+|p/q|^2}}; \delta = arg({{A(D^0\rightarrow K^+ \pi^-)}\over{A{(\overline{D}^0\rightarrow K^+\pi^-)}}})
\end{equation}
\end{itemize}
In these equations $x^{'\pm}$ and $y^{'\pm}$ identify the flavor of the $D$ sample, $i.e.$ $D^0(+)$ $\overline{D}^0(-)$. The mixing parameters $x^{''}$ and $y^{''}$,
\begin{equation}
 x^{''}=x  cos \delta_{K \pi \pi} + y  sin \delta_{K \pi \pi}
\end{equation}
\begin{equation}
 y^{''}=-x  sin \delta_{K \pi \pi} + y  cos \delta_{K \pi \pi}
\end{equation}
may be measured by means of a time-dependent Dalitz plot analysis of the
three-body WS event decay $D^0\rightarrow K^+ \pi^- \pi^0$. The strong phase that rotates
the mixing parameters $x$, $y$ are defined here as $\delta_{k\pi\pi}$=arg($A(D^0\rightarrow K^+ \rho ^-)$/$A(\overline{D}^0\rightarrow K^+ \rho ^-)$)

\section{Mixing}

Charm mixing was established by BABAR and Belle \cite{babar} \cite{belle} using the techniques previously described. 
However the precision
with which the parameters $x_D$ and $y_D$ have been measured can still be improved. A
better precision in the determination of the mixing parameters may allow an improved
understanding of mixing in the up-sector and of whether the CP symmetry is broken in mixing.
In fact the uncertainties on $x_D$ and $y_D$ are of the order of $2 \times 10^{-3}$, which is still too large
to evaluate if there is any CP difference in between $D^0$ and $\overline{D}^0$.
Different measurements of charm mixing may be combined and projected into $(x_D,y_D)$, as shown in Fig.[1]
This would require:
\begin{itemize}
\item $\chi^2$ minimization technique.

\item Correlation effects need to be taken into account.
\subitem  $(x^{'2},y^{'})$ from WS $D^0\rightarrow K^+ \pi^-$ decays.
\subitem  $(x^{''},y^{''})$ from time-dependent Dalitz plot (TDDP) analysis of  $D^0\rightarrow K^+ \pi^- \pi^0$.
\subitem  $y_{CP}$ from tagged/untagged  $D^0\rightarrow h^+ h^-$. 
\subitem  $(x_D,y_D)$ from the combined channels.
\item $CP$ conserving hypothesis.
\end{itemize}
The results of this analysis are given in table [1].
\begin{table}[here!]
\begin{center}
\begin{tabular}{l|cccc}
Fit &  $x \times 10^{-3}$ &  $y \times 10^{-3}$ & $\delta^0_{k^+\pi^-}$  &   $\delta^0_{k^+\pi^-\pi^0}$ \\ 
\hline
   &   $xxx \pm 0.19$     &     $yyy \pm 0.11$      &     $\delta \delta \delta \pm 0.71$  & $\delta \delta \delta \pm 0.83$    \\
\hline
\end{tabular}
\caption{
Mixing parameters $(x_D,y_D)$ and strong phases $\delta_{K\pi}$ and $\delta_{K\pi\pi}$ from a $\chi^2$ fit to observables obtained from $SuperB$ 
when 1.0 $ab^{-1}$ of simulated data at the charm threshold is considered. The
 central value is arbitrarily chosen, and for this reason it is not shown.
}
\label{tab:blood}
\end{center}
\end{table}

It is clear that the $SuperB$ experiment not only will improve our knowledge of
charm mixing using a large data sample collected at the $\Upsilon(4S)$, but with the machine
running at the charm threshold it will be possible to increase the sensitivity to the
mixing parameters. In fact, with a run at charm threshold, one can reduce the size of
$D^0 \rightarrow K_s h^+ h^-$ ellipse. This reduces the area of the WS $D^0 \rightarrow K^+ \pi^-$ ellipse that
combines strong phase measurement and $\Upsilon(4S)$ analysis of time-distribution of $k^+ \pi^-$ decays, and
reduces the area of the WS  $D^0 \rightarrow K^+ \pi^- \pi^0$ decays. Another interesting
aspect of these studies is the possibility to define the $charm$ $golden$ $channels$ $D^0 \rightarrow K_s h^+ h^-$ $(h=\pi,K)$
which are self-conjugate multi-body final states and represent a combination of $CP$ odd and even eigenstates.
If the measurement of strong phases yields $\delta _f =0,\pi$ then the parameters $x_D$ and $y_D$ are directly
mesurable with a TDDP analysis.

\begin{figure}[here!]
\begin{center}
\includegraphics[width=1.0\textwidth]{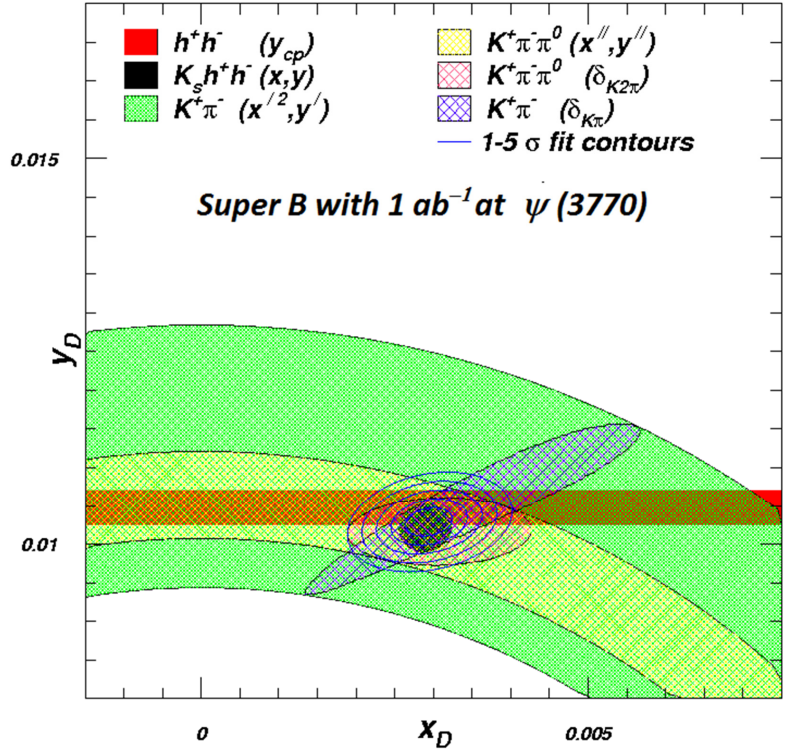}
\caption{Main charm mixing parameters combined into average values for $x_D$ and $y_D$ when considering
         a 1.0 $ab^{-1}$ of data collected at the charm threshold (including projections of strong phase measurements
         $\delta_{k\pi}$ and $\delta{k \pi \pi}$) and 75 $ab^{-1}$ of data collected at the $\Upsilon(4S)$. The contours
         range from 1 to 4 standard deviations two-dimensional confidence regions from the $\chi^2$ fit to these
         results are shown as solid lines }
\label{fig:magnet}
\end{center}
\end{figure}


\section{CP Violation}
$CP$ violation has been discovered in the Kaon system \cite{kaon} long ago and more
recently in the B meson one.
In the charm sector $CP$ violation is expected to be very small.
Moreover the charm sector is the only $up-type$ sector where we can expect
to perform these measurements. 
\subsection {Indirect $CP$ Violation}
As shown in eq. [6] indirect $CP$ violation may be manifest through asymmetries in a broad class of observables. Effective values of the mixing
parameters are defined by measuring separately the $D^0$ and $\overline{D}^0$
mesons. In this case we have:

\begin{equation}
     D^0\rightarrow (x^{+}_D,y^{+}_D),
\end{equation}
\begin{equation}
     \overline{D}^0\rightarrow (x^{-}_D,y^{-}_D),
\end{equation}
where the sign +/- depends on the electric charge of the charm quark ($Q_c$=+2/3 and
$Q_{\overline{c}}$=-2/3). Ignoring systematic uncertainties (which will be almost 
identical for both the $D$ and $\overline{D}^0$ mesons and can be neglected) 
$SuperB$ will be able to measure a difference of the order of 5 $\times 10^{-4}$ 
in $x^+_D$-$x^-_D$ and  3$\times 10^{-4}$ in $y^+_D$-$y^-_D$ at 3$\sigma$ level.
If these differences will be observed and interpreted as being due to
$CP$ violation in mixing, then they would provide a measurement of
$|{q_D\over{p_D}}|$. 
The following asymmetry can furthermore be
measured:

\begin{equation}
     a_z={{(z^{+}-z^{-})}\over{(z^{+}+z^{-})}}\approx {{(1-|{p_D\over{q_D}}|^2)}\over{(1+|{p_D\over{q_D}}|^2)}}
\end{equation}
$z$ can be either $x_D$ or $y_D$. The same study may be applied assuming that
$z$ is $y_{CP}$, $y^{'}$, $x^{''}$, $y^{''}$. 
Estimates of the expected uncertainties that $SuperB$
may obtain by combining different modes are given in Table [2].

\begin{table}[b]
\begin{center}
\begin{tabular}{ccc}  
Parameter &  Mode  &  $\sigma(|q_D/p_D|\times 10^2)$ \\ \hline 
 $x_D$  &   all modes     &     $\pm$1.8  \\
 $y_D$ &  all modes    &      $\pm$1.1 \\ \hline
\end{tabular}
\caption{Combination of estimated uncertainties in the CPV mixing parameter $q_D/p_D$ that may be obtained at $SuperB$ when considering the effective values of
the mixing parameters from 75 $ab^{-1}$ of data collected at the $\Upsilon (4S)$}
\label{tab:blood}
\end{center}
\end{table}

\subsection {Direct $CP$ Violation}
The LHCB Collaboration and CDF recently reported the first hint 
of time integrated $CP$ asymmetry by combining the measurement for
$D^0\rightarrow k^+k^-$, respectively at 3.5 and 2.7 $\sigma$ level \cite{lhcb} \cite{CDF}.
To evaluate this asymmetry the quantity:

\begin{equation}
     A_{CP}(h^+h^-)={{\Gamma(D^0\rightarrow h^+h^-)-\Gamma(\overline{D}^0\rightarrow h^+h^-)}\over{\Gamma(D^0\rightarrow h^+h^-)+\Gamma(\overline{D}^0\rightarrow h^+h^-)}}\approx A^{dir}_{CP}(h^+h^-)+{<t(h^+h^->\over{\tau}}A^{ind}_{CP} 
\end{equation}
has been defined.
New measurements
with better precision are needed in order to understand if this result
is due to new physics. In any case following the same approach the $SuperB$
experiment can reach a sensitivity of the order of $\sigma=3\times 10^{-4}$.

\subsection{Time-dependent $CP$ violation analysis}
In Ref. \cite{tdcpv} time-dependent $CP$ violation $(TDCPV)$ studies have been proposed 
for charm using a very similar formalism to that adopted when studying a $B_d$.
The $SuperB$ experiment will collect data at the charm threshold producing
correlated $D^0$ mesons and at the $\Upsilon(4S)$ producing un-correlated ones.
It will be possible to perform measurements in the two different
configurations of the machine. Observations of $TDCPV$ in charm can
be used to constrain the angle $\beta_{c,eff}$ in the $charm$ $unitarity$ $triangle$ and the time-dependent analysis in general can be used to measure mixing parameters when $TDCPV$ is observed.

\subsection{Sensitivity to $\beta_{c,eff},\phi_{MIX},\phi_{CP}$ and $x_D$}

Expected sensitivities on the parameters described in the previous section 
are reported here for un-correlated $D^0$ mesons production. Further results
are given in \cite{tdcpv}. Table [3] shows the expected uncertainties for
$\phi_{MIX}$ and $\sigma(\beta_{c,eff})$ while Table [4] shows the expected 
sensitivities on $x_D$.

\begin{table}[here!]
\begin{center}
\begin{tabular}{cc}  
Parameter & Un-correlated $D$'s ($\Upsilon(4S)$) \\ \hline 
 $\phi_{\pi \pi}$=${arg(\lambda_{\pi\pi})}$  &   $2.2^o$            \\
 $\phi_{kk}$==${arg(\lambda_{KK})}$=  &   $1.3^o$            \\
 $\beta_{c,eff}$ &   $1.3^o$           \\ \hline
\end{tabular}
\caption{Summary of expected uncertainties from 75 $ab^{-1}$ of data at $\Upsilon(4S)$.}
\label{tab:bloo}
\end{center}
\end{table}
\begin{table}[here!]
\begin{center}
\begin{tabular}{ccc}  
Mode & $\sigma_{x_D}(\phi=\pm 10^o)$ & $\sigma_{x_D}(\phi=\pm 20^o)$ \\ \hline 
 $D^0\rightarrow \pi^+\pi^-$  &   0.12 \% &  0.06 \%          \\
$D^0 \rightarrow k^+k^-$  &   0.08 \% & 0.04 \%           \\ \hline
\end{tabular}
\caption{Estimates of the sensitivity on $x_D$ when 75$ab^{-1}$ of data collected at the $\Upsilon(4S)$ is considered
for the decays $D^0\rightarrow \pi^+\pi^-$ and $D^0\rightarrow K^+K^-$ and $\phi= \phi_{MIX}-2\beta_{c,eff}$}
\label{tab:bloo1}
\end{center}
\end{table}
This shows that a time-dependent analysis applied to charm not only may help us
to better understand the flavor changing structure of the SM, but the observation
of $CP$ violation in charm may allow us to study the charm triangle through the
measurement of the angle $\beta_{c,eff}$. This will provide us with a consistency check of
the CKM mechanism. It is important to mention that with current experimental
sensitivity, any observation of a value of the $\beta_{c,eff}$ inconsistent with zero will be a clear signal of new physics.
On the other hand, the time-dependent analysis can measure $x_D$ and $\phi_{MIX}$ 
using not only the decay modes $D^0 \rightarrow K^+ K^-$ or $D^0 \rightarrow \pi^+ \pi^-$,
but additional decay modes are available to carry out time-dependent mixing-related
measurements. Decay channels including neutral states, as it is in the
case of $D^0 \rightarrow K^0_s \pi^0$, 
with  branching ratios that are larger than those for $D^0 \rightarrow K^+ K^-$ and $D^0 \rightarrow \pi^+ \pi^-$ may provide better constraints on the determination of mixing phase/parameters,
and this makes an electron-positron machine unique when
performing such a measurement.

\section{Conclusions}

The SuperB experiment with its run at the charm
threshold will be able to perform strong phase measurements that will shrink the
allowed parameter space, and it has been shown that by using the $effective$ $values$ for the
mixing parameters it is possible to perform a test of $CP$ violation (indirect).
Direct CP violation has been discussed, and it has been highlighted that it is not
yet clear if the Standard Model may account for the current experimental central value, or if some
new physics is showing up, and studies of additional modes help resolve this question.
$SuperB$ may play a leading role in the understanding of direct $CP$ violation repeating
the analysis already carried out, and performing new ones that would otherwise be difficult
 for other experiments (final states including neutrals).
Finally, time-dependent CP violation has been introduced and it was shown that
with that formalism it is possible to perform measurements of the quantities: $\phi_{MIX}$,
$x_D$, and $\beta_{c,eff}$.

\end{document}